\documentclass[twocolumn,showpacs,preprintnumbers,amsmath,amssymb,prb]{revtex4}

\usepackage{graphicx}
\usepackage{dcolumn}
\usepackage{bm}

\begin{document}

\title{Optically tuned dimensionality crossover in photocarrier-doped SrTiO$_3$: onset of weak localization}

\author{Y. Kozuka}
 \email{kk56117@mail.ecc.u-tokyo.ac.jp}
\author{Y. Hikita}
\author{T. Susaki}
\affiliation{Department of Advanced Materials Science, University of Tokyo, Kashiwa, Chiba 277-8561, Japan}
\author{H. Y. Hwang}
\affiliation{Department of Advanced Materials Science, University of Tokyo, Kashiwa, Chiba 277-8561, Japan}
\affiliation{Japan Science and Technology Agency, Kawaguchi, 332-0012, Japan}

\begin{abstract}
We report magnetotransport properties of photogenerated electrons in undoped SrTiO$_3$ single crystals under ultraviolet illumination down to 2 K. By tuning the light intensity, the steady state carrier density can be controlled, while tuning the wavelength controls the effective electronic thickness by modulating the optical penetration depth. At short wavelengths, when the sheet conductance is close to the two-dimensional Mott minimum conductivity we have observed critical behavior characteristic of weak localization. Negative magnetoresistance at low magnetic field is highly anisotropic, indicating quasi-two-dimensional electronic transport. The high mobility of photogenerated electrons in SrTiO$_3$ allows continuous tuning of the effective electronic dimensionality by photoexcitation.
\end{abstract}

\pacs{73.20.Fz, 72.40.+w, 72.20.Fr, 77.84.Bw}

%73.20.Fz 	Weak or Anderson localization
%72.40.+w 	Photoconduction and photovoltaic effects
%72.20.Fr 	Low-field transport and mobility; piezoresistance
%77.84.Bw 	Elements, oxides, nitrides, borides, carbides, chalcogenides, etc.

\maketitle

\section{INTRODUCTION}
Metal-insulator (MI) transitions have been a longstanding problem in solid state physics, ranging from conventional semiconductors to strongly correlated materials.\cite{Imada} The disorder-induced MI transition was examined for noninteracting electronic systems by Abrahams \textit{et al}., indicating that there is a metal-insulator transition for $d>2$ (where $d$ is the dimension), while all states are localized for $d\le 2$ due to interference between coherently scattered electrons.\cite{Abrahams1} However, metallic behavior has been observed in a number of two-dimensional systems and the importance of electron-electron interactions has been suggested.\cite{Abrahams2} Even in such case, the Ioffe--Regel (IR) criterion\cite{Ioffe} $l_{\mathrm{min}}\simeq a$ gives a satisfactory estimate of the minimum metallic conductivity of $e^2/h$ in two dimensions, where $l_{\mathrm{min}}$ is the minimum mean free path, $a$ is the spacing of neighboring atoms, $e$ is the elementary electric charge, and $h$ is the Planck constant.\cite{Abrahams2} In three-dimensional systems, resistivity saturation has been observed around the conductivity given by the IR criterion $2e^2k_{F}/3\pi h$ ($k_{F}$: Fermi wavenumber) in a wide range of metals, including heavy-fermion systems.\cite{Hussey} In spite of the apparent universality, violation of the IR criterion is common in many transition-metal oxides, keeping a metallic temperature dependence over the IR limit.\cite{Hussey}\par
In this context, SrTiO$_3$ is an interesting material, bridging the gap between conventional semiconductors and strongly correlated transition-metal oxides. This is  because doped SrTiO$_3$ exhibits extremely high electron mobility exceeding 10$^4$ cm$^2$ V$^{-1}$ s$^{-1}$,\cite{Tufte} despite the $d$-band conduction with a large effective mass ($m^{\ast}$), similar to strongly correlated materials. Owing to the exceptionally large dielectric constant,\cite{Sawaguchi} the Bohr radius ($a_{H}$) is expected to be much larger than the lattice constant of SrTiO$_3$. Consequently, the transport properties of SrTiO$_3$ have been understood within a conventional description based on the relaxation time approximation.\cite{Tufte} In addition, the observed MI transition\cite{Lee} around the critical carrier density ($n_{c}$) of 10$^{18}$ cm$^{-3}$ has been also interpreted in terms of the Mott criterion\cite{Mott} $a_{H}n_{c}^{1/3}\simeq 0.25$ for oxygen-deficient SrTiO$_{3-\delta}$ using $m^{\ast}=12m_{0}$ ($m_{0}$: bare electron mass) and the room-temperature value of $\epsilon=220\epsilon_{0}$,\cite{Cox} where $a_{H}$ is defined by $\epsilon \hbar^{2}/m^{\ast}e^{2}$ ($\hbar=h/2\pi$), $\epsilon$ is the dielectric constant of the medium, and $\epsilon_{0}$ is the vacuum dielectric constant. Although the Mott criterion gives satisfactory estimates of $n_{c}$ for a great variety of semiconductors,\cite{Edwards} the above discussion for SrTiO$_3$ is questionable because of the highly temperature dependent dielectric constant\cite{Sawaguchi} and because of the absence of any insulating behavior in Nb-doped SrTiO$_{3}$ crystals down to $n\sim 10^{17}$ cm$^{-3}$.\cite{Tufte} Furthermore, recent studies suggest that oxygen vacancies in SrTiO$_{3}$ cannot be treated as random point defects, but are rather highly clustered around extended crystalline defects.\cite{Szot} The resulting inhomogeneity in doping obscures the intrinsic MI transition. In order to extract intrinsic properties of the MI transition, it is crucial to observe critical behavior of the MI transition, namely weak localization, which has not been reported in SrTiO$_3$.\par
Here we report the investigation of transport properties of undoped SrTiO$_3$ single crystals under ultraviolet illumination down to 2 K. Photogenerated electrons exhibit high electron mobility comparable to bulk chemically doped samples, while holes do not contribute to transport. By tuning the wavelength of illumination to shorter wavelength, the electronic thickness is greatly reduced by the reduced optical penetration depth. In this regime we have observed weakly localized behavior such as negative magnetoresistance when the sheet conductance is close to the two-dimensional Mott minimum conductivity. Photocarrier doping may be one of the most suitable techniques to characterize such critical behavior because of the immediate modulation of carrier density just by adjusting the intensity of illumination.\cite{Katsumoto} In addition, the effect of inhomogeneity can be minimized as long as the light intensity profile is uniform on the sample surface. Magnetoresistance measurements in different geometries revealed that the electron conduction tends to be two-dimensional at short wavelength. The result of longitudinal magnetoresistance measurement indicated that the dominant conduction occurs within the thickness estimated from the absorption coefficient of SrTiO$_3$. We have also confirmed disappearance of negative magnetoresistance when the light intensity is increased so that the zero-field conductivity becomes much larger than $e^2/h$ or when the wavelength of the illumination is near the absorption edge, giving a large penetration depth. The observed negative magnetoresistance is much smaller than theoretically predicted, which may be understood by taking into account competing effects such as disorder and electron-electron interactions.
\section{EXPERIMENTAL METHODS}
We have employed an optical system (Asahi Spectra Co.) equipped with a 100 W Xe lamp as a light source. The irradiated light is attenuated by neutral density (ND) filters, monochromatized by a band-pass filter (10 nm bandwidth), transmitted through a fiber bundle into a superconducting magnet dewar, and eventually projected on samples by imaging the output of a rod lens in an square shape of $\simeq2\times2$ cm$^2$. The maximum intensity is $\sim1$ mW/cm$^2$ and the intensity can be almost continuously reduced down to 0.001 \% by the ND filters. Magnetoresistance and Hall measurements were performed by a standard four-terminal dc technique with electrical contacts made by Al wire bonding. The resistance was measured in the range of low electric field to avoid previously reported nonohmic effects.\cite{Grabner}. The lowest attainable temperature was 2 K under full illumination.\par
Commercial SrTiO$_3$ single crystals were used (Furuuchi Chemical Co. and Shinkosha Co.), which were supplied polished to an optical grade. Typical sample dimensions are $3.0\times0.5\times0.5$ mm$^{3}$, where the inhomogeneity of the light intensity is less than 1\% across the sample surface. In the temperature range used in this study, no significant sample dependence was observed in the normalized transport properties. The absolute magnitude, however, varied over a range of a factor of two in sample-to-sample variations. In the subsequent results, the temperature was increased above 150 K before changing the band-pass filters in order to remove any history effects originating from trapping states.\cite{Amodei} In addition, the samples were exposed to illumination for an extended period of time to reach steady state before measurements.
\section{RESULTS AND DISCUSSIONS}
\subsection{Basic photoconducting properties}
Electron-hole pairs are expected to be generated in SrTiO$_{3}$ across the band gap between the O$_{2p}$ bands and the Ti$_{3d}$ bands in response to ultraviolet irradiation. Hall measurements indicated that the conduction is dominated by electrons as reported previously.\cite{Yasunaga} This fact illustrates that the mobility of holes is much lower than that of electrons,\cite{Tufte,Kholkin} or that holes are immediately trapped after generation.\cite{Feng} Although light illumination generates nonequilibrium carriers, conventional descriptions are valid when thermal equilibrium is reached much faster than recombination. This is justified in the subsequent measurements owing to the extremely long carrier lifetime, which approaches 1 ms below 30 K\cite{Hasegawa} if we take photoluminescence decay as a measure of the lifetime. Figure 1 shows typical transport data for illumination at 370 nm, just above the band gap of SrTiO$_{3}$. The sheet conductance ($\sigma_{\mathrm{2D}}$) increases with decreasing temperature, as does the sheet carrier density ($n_{\mathrm{2D}}$) to a lesser degree. Varying the light intensity tunes the steady state carrier density, and the deduced Hall mobility ($\mu$) increases with decreasing temperature, indicating metallic behavior.\par
\begin{figure}[tbp]
  \begin{center}
    \includegraphics{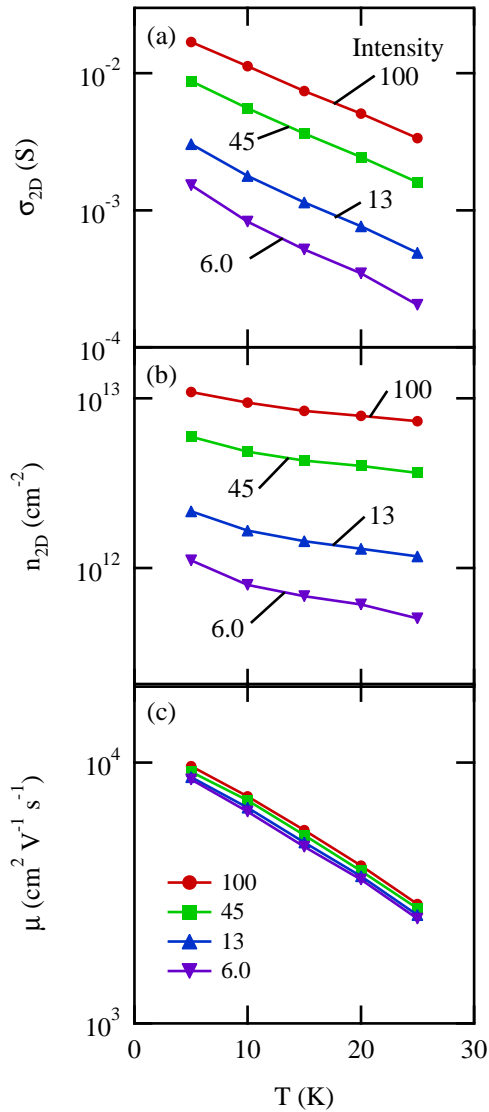}
  \end{center}
  \caption{(Color online) Temperature dependence of (a) sheet conductance, (b) sheet carrier density, and (c) Hall mobility for varying light intensity at 370 nm. The values indicated are relative light intensity (100: $\sim1$ mW/cm$^{2}$).}
\end{figure}
\begin{figure}[tbp]
  \begin{center}
    \includegraphics{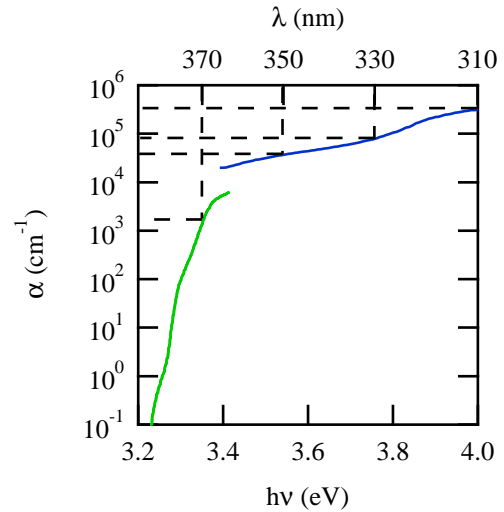}
  \end{center}
  \caption{(Color online) Absorption coefficient of SrTiO$_3$ as a function of energy of incident light. The low-energy curve is taken from Ref. \onlinecite{Capizzi}, while the high-energy curve is taken from Ref. \onlinecite{Cohen}.}
\end{figure}
\begin{figure}[tbp]
  \begin{center}
    \includegraphics{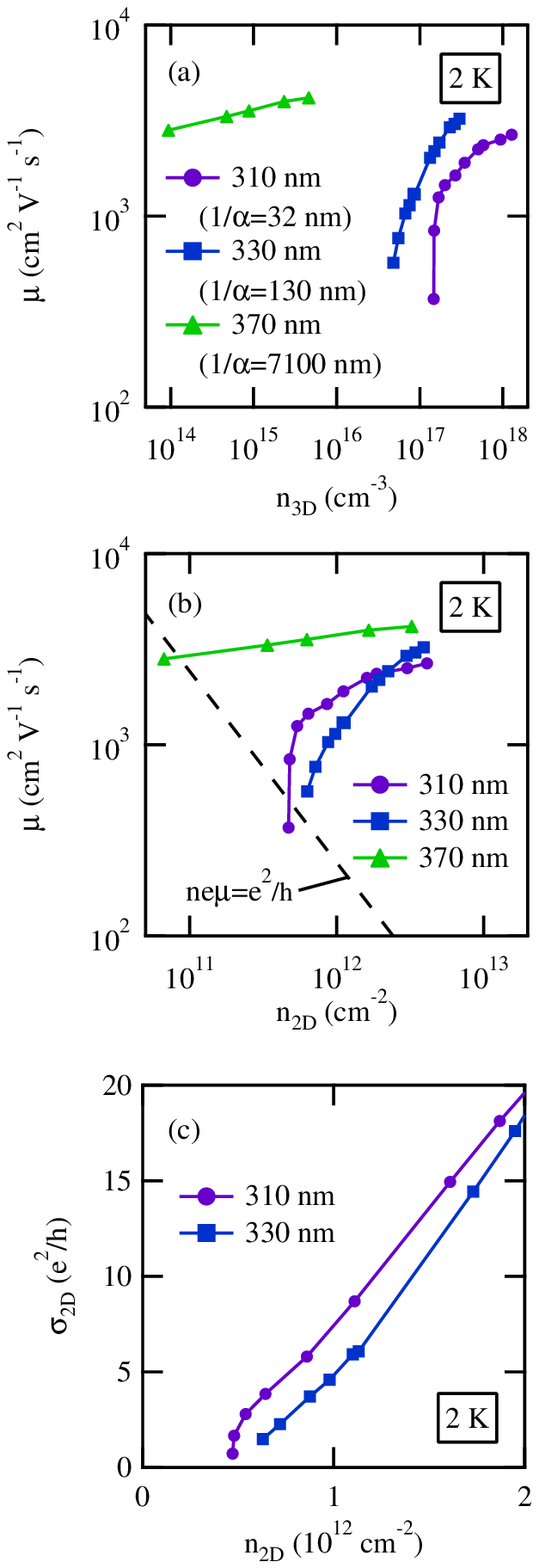}
  \end{center}
  \caption{(color online) Hall mobility as a function of (a) $n_{\mathrm{3D}}$ and (b) $n_{\mathrm{2D}}$ with varying intensity and wavelength of illumination. The thickness of the conducting layer is approximated by $1/\alpha$ in (a). The dashed line in (b) indicates the two-dimensional Mott minimum conductivity $n_{\mathrm{2D}}e\mu=e^{2}/h$. (c) Sheet conductance as a function of sheet carrier density near the mobility drop in (b). The measurement temperature for all panels is 2 K, and the carrier density was controlled by changing the light intensity.}
\end{figure}
The depth profile of the electron distribution can be estimated from the optical penetration depth, deduced from the absorption coefficient ($\alpha$) of SrTiO$_{3}$ as shown in Fig. 2. Assuming electron diffusion can be neglected (reasonable given the lack of hole mobility in SrTiO$_{3}$), the thickness of the conducting layer is approximated by $1/\alpha$: $1/\alpha\simeq 7.1$ $\mu$m at a wavelength $\lambda=370$ nm.\cite{Capizzi} By going to higher energy, this thickness can be reduced by orders of magnitude ($1/\alpha\simeq 32$ nm for $\lambda=310$ nm).\cite{Cohen} Figure 3(a) shows the variation of $\mu$(2 K) as a function of the three-dimensional carrier density ($n_{\mathrm{3D}}$) for varying $\lambda$, where $1/\alpha$ was used to estimate $n_{\mathrm{3D}}$. The behavior is opposite that expected in doped semiconductors, in that the lowest densities reached using $\lambda=370$ nm show higher $\mu$ than for much higher density data taken at shorter $\lambda$. In this low density regime, $\mu$ typically decreases with decreasing $n_{\mathrm{3D}}$ due to reduced electron-electron screening. These results can be understood in terms of a dimensionality crossover with decreasing optical penetration depth. Figure 3(b) replots the data of Fig. 3(a) in terms of $n_{\mathrm{2D}}$. The sharp reduction in $\mu$ for $\lambda=330$ nm and $\lambda=310$ nm occurs in the vicinity of the two-dimensional Mott minimum conductivity, indicative of a metal-insulator transition. For $\lambda=370$ nm, by contrast, this feature is not observed, indicating that the two-dimensional conductivity threshold is not relevant at this wavelength.\par
We investigated the behavior of the conductivity near the mobility drop as shown in Fig. 3(c). Although this analysis should be carried out for the $T=0$ extrapolated values of the conductivity, in the absence of lower-temperature data, we use the conductivity at 2 K for a semiquantitative analysis. The conductivity seems to follow $\sigma=\sigma_{c}\left[(n/n_{c})-1\right]^{\zeta}$, and the critical exponent $\zeta$ seems globally unity as in compensated semiconductors,\cite{Rosenbaum} where $\sigma_{c}$ is a constant. However, lower-temperature measurements near $n_{c}$ are necessary for a rigorous determination. For $\lambda=370$ nm, we have not observed three-dimensional critical behavior which indicates a MI transition. According to the Mott criterion, $n_{c}$ of SrTiO$_3$ is $4.5\times10^{13}$ cm$^{-3}$ using $m^{\ast}=1.2m_{0}$\cite{Uwe} and a low-temperature value of $\epsilon=20000\epsilon_{0}$,\cite{Sawaguchi} which corresponds to a Fermi temperature ($T_{F}$) of 45 mK. Therefore, based on this estimation, our measurement temperature far exceeds $T_{F}$, and the system may be regarded as a nondegenerate semiconductor.
\subsection{Temperature dependent mobility correction}
The characteristic temperature dependent conductivity for the case of two-dimensional weak localization is given by
\begin{equation}
\Delta\sigma(T)=\sigma(T)-\sigma_{0}=\frac{pe^2}{\pi h}\ln\left(\frac{T}{T_0}\right),\label{eq:2DWL}
\end{equation}
where $\sigma_{0}$ is a constant, $p$ is a constant satisfying $1/\tau_{\epsilon}\propto T^{p}$, $\tau_{\epsilon}$ is the inelastic relaxation time, and $T_0$ is a constant.\cite{Abrahams1} In the framework of the theory of weak localization, carrier density is constant and the conductivity correction stems from the variation of mobility as a function of temperature, when the Hall mobility is expressed as $\mu=1/\rho_{0}ne$ ($\rho_{0}$: zero-field resistivity).\cite{Altshuler1}  In our measurement, carrier density also varies with temperature because of the variation of photocarrier generation efficiency at a constant light intensity, as shown in Fig. 1(b).\par
\begin{figure}[tbp]
  \begin{center}
    \includegraphics{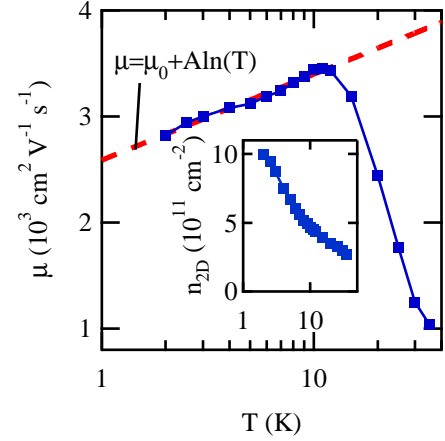}
  \end{center}
  \caption{(Color online) Hall mobility as a function of temperature under 330 nm illumination for a sample $n_{\mathrm{2D}}$(2 K)=$1.0\times10^{12}$ cm$^{-2}$ and $\mu$(2 K)=2800 cm$^2$ V$^{-1}$ s$^{-1}$. The dashed line is the result of fitting by the equation $\mu(T)=\mu_0+A\ln(T)$, where $\mu_0=2600$ cm$^2$ V$^{-1}$ s$^{-1}$ and $A=350$ cm$^2$ V$^{-1}$ s$^{-1}$. The inset shows $n_{\mathrm{2D}}(T)$ with a constant light intensity.}
  \label{fig:Temperature.eps}
\end{figure}
Figure 4 shows the temperature dependence of $\mu$ and $n_{\mathrm{2D}}$ for a sample illuminated at $\lambda=330$ nm. The mobility exhibits a logarithmic temperature dependence, which is consistent with Eq. (1). For a rough quantitative estimation, we fix $n_{\mathrm{2D}}=5\times10^{11}$ cm$^{-2}$; fitting the data in Fig. 4 by the equation $\mu(T)=\mu_0+A\ln(T)$ leads to $neA=2.8\times10^{-5}$ S, which is in good agreement with the theoretically predicted value $e^2/\pi h=1.2\times10^{-5}$ S. The fact that $p\simeq 2$ may imply that the dominant inelastic scattering mechanism is an electron-electron correlation.
\subsection{Magnetoresistance}
Another characteristic response of weak localization appears in the low-field magnetoresistance. Figure 5 shows the transverse magnetoresistance under varying intensity and wavelength of  illumination at 2 K. Under 310 nm illumination, we observed a small negative magnetoresistance at low field, which is typical of weak localization. This feature diminishes with increasing light intensity and increasing $\lambda$, until by $\lambda=370$ nm no negative magnetoresistance is observed. The large positive magnetoresistance under high magnetic field originates from a cyclotron motion driven by the Lorentz force, following $\Delta R(H)/R(0) = \xi(\mu H)^2$, where $\xi$ is a constant of the order of unity, related to the details of the Fermi surface and the shape of the sample.\cite{Schroder} This wavelength dependent magnetoresistance follows the dimensionality crossover indicated in Fig. 3. Under shorter wavelength illumination, electrons tend to localize with decreasing carrier density because the small penetration depth enhances their two-dimensional nature. Under longer wavelength illumination, electrons extend more three-dimensionally, and the reduced localization effect would be smeared out by the strong positive magnetoresistance.\par
\begin{figure}[tbp]
  \begin{center}
    \includegraphics{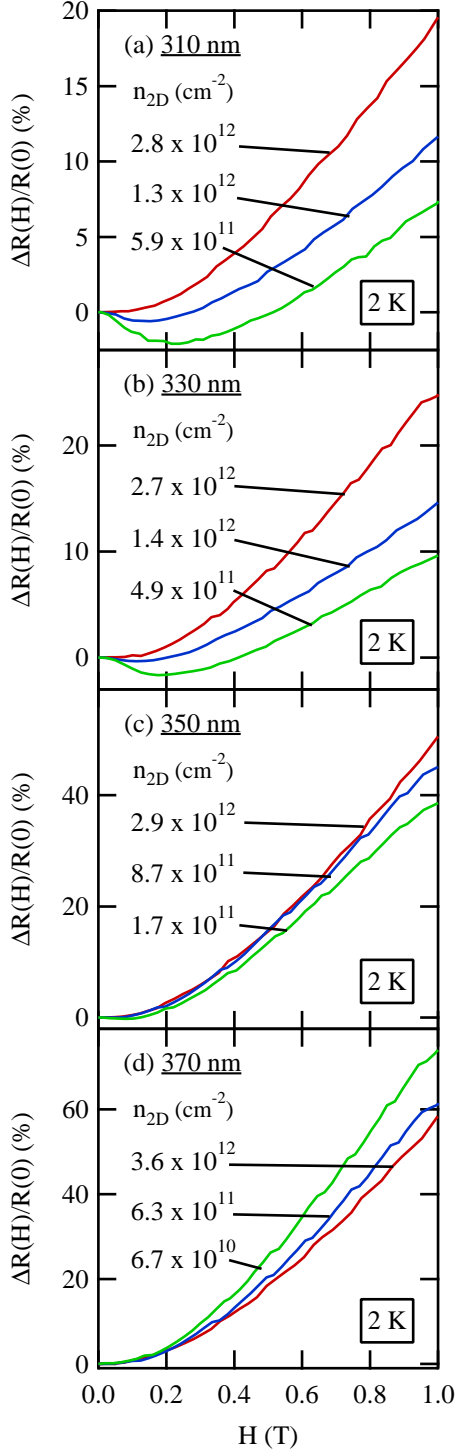}
  \end{center}
  \caption{(Color online) Transverse magnetoresistance under (a) 310 nm, (b) 330 nm, (c) 350 nm, and (d) 370 nm illumination with varying light intensity at 2 K.}
\end{figure}
Another reason of the absence of the negative magnetoresistance for $\lambda=370$ nm may be the rather small $T_{F}$. This is because weak localization can be applied only in a degenerate regime, namely $T<T_{F}$. We estimate the Fermi energy in the cases of 310 nm and 370 nm. A typical $n_{\mathrm{2D}}$ of $5\times10^{11}$ cm$^{-2}$ corresponds to $n_{\mathrm{3D}}=2\times10^{17}$ cm$^{-3}$ for $\lambda=310$ nm, and $n_{\mathrm{3D}}=7\times10^{14}$ cm$^{-3}$ for $\lambda=370$ nm. Consequently, $T_{F}$ is estimated as 10 K for $\lambda=310$ nm and 0.3 K for $\lambda=370$ nm within a free carrier approximation with an effective mass of 1.2$m_{0}$.\cite{Uwe} As previously, we conclude our measurements at this $\lambda$ are in the nondegenerate regime.\par
\begin{figure}[tbp]
  \begin{center}
    \includegraphics{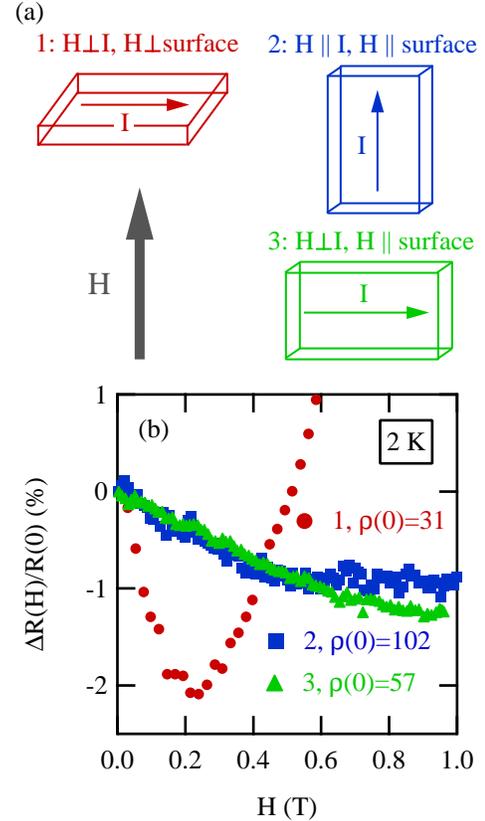}
  \end{center}
  \caption{(Color online) (a) Schematic diagram of sample geometries with respect to the directions of current and magnetic field. 1: $H\perp I$ and $H\perp$ sample surface, 2: $H\parallel I$ and $H\parallel$ sample surface, and 3: $H\perp I$ and $H\parallel$ sample surface. (b) Magnetoresistance in the three geometries under 310 nm illumination at 2 K. $\rho(0)$ gives the sheet resistance at $H=0$ in the units of k$\Omega/\Box$. The sample geometries numbered in (a) correspond to the data points numbered in (b).}
\end{figure}
\begin{figure}[tbp]
  \begin{center}
    \includegraphics{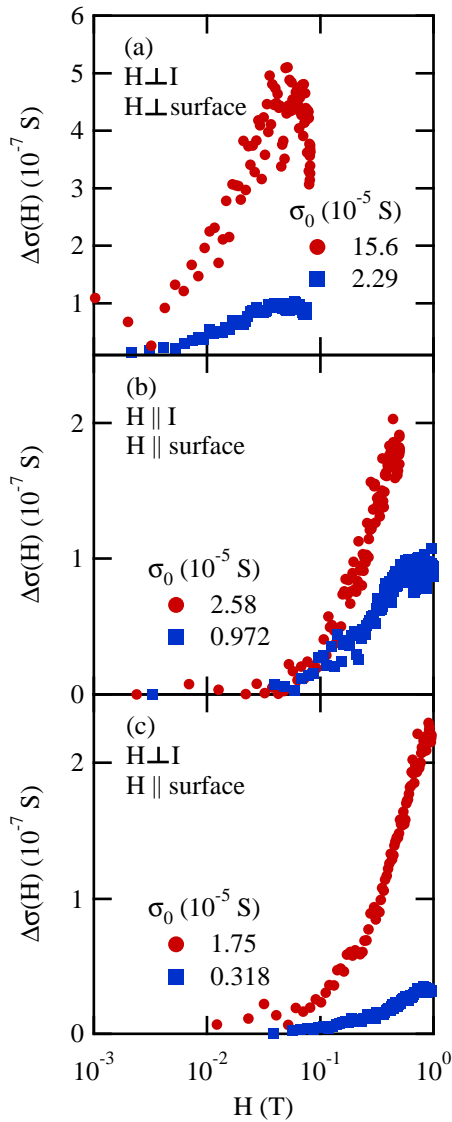}
  \end{center}
  \caption{(Color online) Conductivity correction by magnetic field (a) in geometry 1, (b) in geometry 2, and (c) in geometry 3 as shown in Fig. 6 (a) with varying light intensity. $\sigma_{0}$ gives the sheet conductance at $H=0$.}
\end{figure}
Two-dimensionality of the electron gas under shorter wavelength illumination is further supported by a geometrical dependent magnetoresistance as shown in Fig. 6, where the geometries numbered in (a) correspond to those in (b). As shown in this figure, the magnetoresistance exhibits anisotropy, indicating that the magnetoresistance is caused by an orbital effect. Moreover, the magnetoresistance in geometry 2 resembles that in geometry 3. This clearly reflects two-dimensional behavior because the magnetoresistance in geometry 3 should be identical to the case of geometry 1 if the electronic system were completely three-dimensional. However, the longitudinal magnetoresistance should be absent in a purely two-dimensional system. In fact, similar anisotropic magnetoresistance has been observed in metallic thin films such as In$_2$O$_3$.\cite{Ovadyahu} In analogy with those results, magnetoresistance can be observed in the longitudinal geometry when the radius of the lowest Landau level $L_{H}=\sqrt{\hbar/eH}$ crosses the effective thickness of the conducting layer as the magnetic field increases. In Fig. 6(b), the onset of longitudinal magnetoresistance is about 0.06 T, which corresponds to $L_{H}=105$ nm. This value is comparable to the optical penetration depth value $1/\alpha$($\lambda=310$ nm)=32 nm.\par
The magnetoconductance for weak localization is expressed as
\begin{subequations}\label{eq:AMR}
\begin{equation}
\Delta\sigma_{\perp}(H)=\frac{e^2}{\pi h}\ln\left[\frac{2L_{\epsilon}^{2}}{L_{H}^{2}}\right]\qquad(L_{H}\ll L_{\epsilon})
\end{equation}
for perpendicular magnetic field\cite{Hikami} and
\begin{equation}
\Delta\sigma_{\parallel}(H)=\frac{e^2}{\pi h}\ln\left[\frac{t^{2}L_{\epsilon}^{2}}{12L_{H}^{4}}\right]\qquad(L_{H}^{2}\ll tL_{\epsilon})
\end{equation}
\end{subequations}
for parallel magnetic field,\cite{Altshuler2} where $t$ is the thickness, $L_{\epsilon}=\sqrt{2D\tau_{\epsilon}}$ is the inelastic diffusion length, and $D$ is the diffusion constant. Figure 7 shows magnetoresistance with varying $n_{\mathrm{2D}}$ in the three geometries shown in Fig. 6(a). We observed a logarithmic magnetic field dependence in all three geometries, which is qualitatively consistent with Eqs. (\ref{eq:AMR}). The observed conductivity correction is, however, the order of $10^{-7}$ S even though the light intensity (and thus $n_{\mathrm{2D}}$) is varied, while the theoretically predicted value is the order of $10^{-5}$ S.\par
In considering this discrepancy, we note that we have observed a temperature dependence of mobility roughly consistent in magnitude with weak localization, while the magnetoresistance was about two orders of magnitude smaller than theoretically predicted. This asymmetry suggests another contribution to magnetoresistance. Conventionally, electron-electron interaction is known to give a conductivity correction dependent on temperature and magnetic field, qualitatively similar to weak localization in the range of $k_{F}l\gg 1$ ($l$: mean free path).\cite{Altshuler1} The sign and strength of magnetoresistance from this effect is dependent on the sign (attractive or repulsive) and strength of the electron-electron interaction.\cite{Altshuler3} In a well screened system satisfying $k_{F}/\kappa \gg 1$ ($\kappa$: inverse screening length), magnetoresistance is expected to be small, while the temperature dependent conductivity correction is the same order as Eq. (1). This is true of the present case because of large dielectric constant $\simeq 20000\epsilon_{0}$ of SrTiO$_{3}$ --- the low $T_{F}$ allows dynamic screening by the lattice. On the other hand, we observed a negative magnetoresistance only in the range of $k_{F}l \sim 1$. In addition to the sharp drop of mobility in Fig. 3, these facts indicate that our observations are affected both by weak localization and by electron-electron correlations. 
\section{CONCLUSIONS}
In this study, we have characterized magnetotransport properties of photocarrier-doped SrTiO$_3$ single crystals by changing the intensity and wavelength of illumination. At short wavelength, a drop of the Hall mobility has been observed when the sheet conductance is close to the two-dimensional Mott minimum conductivity, suggesting a two-dimensional MI transition. The uniformity of the light intensity and the tunability of carrier density of photocarrier doping have enabled us to investigate such critical behavior in SrTiO$_3$. Although we have observed characteristic responses of weak localization, the negative magnetoresistance was much smaller than theoretically predicted. Nevertheless, the geometrical dependence of the negative magnetoresistance clearly indicated dominant two-dimensional electronic transport with finite thickness. The discrepancy with weak localization theory may be due to the effect of an electron-electron correlation.\par
An alternative way to precisely control carrier density is electrostatic doping in a field-effect transistor (FET).\cite{Ahn} Recently, it has been successfully applied to SrTiO$_3$, maintaining rather high mobility.\cite{Nakamura} Although an FET is also suitable to confine electrons in two dimensions, the confinement potential may suffer an unfavorable deformation by the gate bias due to the considerably electric-field dependent permittivity of SrTiO$_3$.\cite{Sawaguchi} In contrast, we have demonstrated that the photocarrier doping can be used to precisely tune not only carrier density and but also the effective electronic thickness by separately adjusting the intensity and wavelength of illumination.

\section*{ACKNOWLEDGMENTS}
We thank N. Nagaosa, M. Onoda, H. Takagi, M. Kawasaki, K. S. Takahashi, X. Gao, and S. Okamoto for helpful discussions. This work was partly supported by a Grant-in-Aid for Scientific Research (B) from the Ministry of Education, Culture, Sports, Science and Technology.


\begin{thebibliography}{99}

\bibitem{Imada}M. Imada, A. Fujimori, and Y. Tokura, Rev. Mod. Phys. \textbf{70}, 1039 (1998).

\bibitem{Abrahams1}E. Abrahams, P. W. Anderson, D. C. Licciardello, and T. V. Ramakrishnan, Phys. Rev. Lett. \textbf{42}, 673 (1979).

\bibitem{Abrahams2}E. Abrahams, S. V. Kravchenko, and M. P. Sarachik, Rev. Mod. Phys. \textbf{73}, 251 (2001).

\bibitem{Ioffe}A. F. Ioffe and A. R. Regel, Prog. Semicond. \textbf{4}, 237 (1960).

\bibitem{Hussey}N. E. Hussey, K. Takenaka, and H. Takagi, Phil. Mag. \textbf{84}, 2847 (2004).

\bibitem{Tufte}O. N. Tufte and P. W. Chapman, Phys. Rev. \textbf{155}, 796 (1967).

\bibitem{Sawaguchi}E. Sawaguchi, A. Kikuchi, and Y. Kodera, J. Phys. Soc. Jpn. \textbf{17}, 1666 (1962).

\bibitem{Lee}C. Lee, J. Yahia, and J. L. Brebner, Phys. Rev. B \textbf{3}, 2525 (1971).

\bibitem{Mott}N. F. Mott, \textit{Metal-insulator transitions}, 2nd ed. (Taylor \& Francis, London, 1990).  

\bibitem{Cox}P. A. Cox, \textit{Transition metal oxides: an introduction to their electronic structure and properties} (Clarendon Press, Oxford, 1995), Chap. 4.

\bibitem{Edwards}P. P. Edwards and M. J. Sienko, Phys. Rev. B \textbf{17}, 2575 (1978).

\bibitem{Szot}K. Szot, W. Speier, R. Carius, U. Zastrow, and W. Beyer, Phys. Rev. Lett. \textbf{88}, 075508 (2002).

\bibitem{Katsumoto}S. Katsumoto, F. Komori, N. Sano, and S. Kobayashi, J. Phys. Soc. Jpn. \textbf{56}, 2259 (1987).

\bibitem{Grabner}L. H. Grabner in \textit{Interaction of radiation with solids: proceedings of the Cairo solid state conference at the American University, Cairo, 1966}, edited by A. Bishay (Plenum Press, New York, 1967), p. 155.

\bibitem{Amodei}J. J. Amodei and W. R. Roach in \textit{Proceedings of the third international conference on photoconductivity, Stanford, 1969}, edited by E. M. Pell (Pergamon Press, Oxford, 1971), p. 93.

\bibitem{Yasunaga}H. Yasunaga, J. Phys. Soc. Jpn. \textbf{24}, 1035 (1968).

\bibitem{Kholkin}A. L. Kholkin, V. A. Trepakov, G. A. Smolensky, Ju. V. Likholetov, V. D. Belyaev, and J. Auleutner, Ferroelectrics \textbf{43}, 195 (1982).

\bibitem{Feng}T. Feng, Phys. Rev. B \textbf{25}, 627 (1982).

\bibitem{Hasegawa}T. Hasegawa, M. Shirai, and K. Tanaka, J. Lumin. \textbf{87-89}, 1217 (2000).

\bibitem{Capizzi}M. Capizzi and A. Frova, Phys. Rev. Lett. \textbf{25}, 1298 (1970).

\bibitem{Cohen}M. I. Cohen and R. F. Blunt, Phys. Rev. \textbf{168}, 929 (1968).

\bibitem{Rosenbaum}T. F. Rosenbaum, R. F. Milligan, M. A. Paalanen, G. A. Thomas, R. N. Bhatt, and W. Lin, Phys. Rev. B \textbf{27}, 7509 (1983).

\bibitem{Uwe}H. Uwe, R. Yoshizaki, T. Sakudo, A. Izumi, and T. Uzumaki, Jpn. J. Appl. Phys. \textbf{Suppl.24-2}, 335 (1985).

\bibitem{Altshuler1}B. L. Altshuler, D. Khmel'nitzkii, A. I. Larkin, and P. A. Lee, Phys. Rev. B \textbf{22}, 5142 (1980).

\bibitem{Schroder}D. K. Schroder, \textit{Semiconductor material and device characterization}, 2nd ed. (John Wiley \& Sons, New York, 1998), Chap. 8.

\bibitem{Ovadyahu}Z. Ovadyahu, Y. Gefen, and Y. Imry, Phys. Rev. B \textbf{32}, 781 (1985).

\bibitem{Hikami}S. Hikami, A. I. Larkin, and Y. Nagaoka, Prog. Theor. Phys. \textbf{63}, 707 (1980).

\bibitem{Altshuler2}B. L. Al'tshuler and A. G. Aronov, Pis'ma Zh. Eksp. Teor. Fiz. \textbf{33}, 515 (1981) [JETP Lett. \textbf{33}, 499 (1981)].

\bibitem{Altshuler3}B. L. Al'tshuler, A. G. Aronov, A. I. Larkin, and D. E. Khmel'nitski\v{i}, Zh. Eksp. Teor. Fiz. \textbf{81}, 768 (1981) [Sov. Phys. JETP \textbf{54}, 499 (1981)].

\bibitem{Ahn}C. H. Ahn, J.-M. Triscone, and J. Mannhart, Nature (London) \textbf{424}, 1015 (2003).

\bibitem{Nakamura}H. Nakamura, H. Takagi, I. H. Inoue, Y. Takahashi, T. Hasegawa, and Y. Tokura, Appl. Phys. Lett. \textbf{89}, 133504 (2006).

\end{thebibliography}
\end{document}